\documentclass[a4paper]{article}

\usepackage[english]{babel}
\usepackage[utf8]{inputenc}
\usepackage[colorinlistoftodos]{todonotes}
\usepackage{amsmath}
\usepackage{amssymb}
\usepackage{amsthm}
\usepackage{mathtools}
\usepackage{mathrsfs}
\usepackage{siunitx}
\usepackage{algpseudocode}

\usepackage{tikz}

\usepackage{tabularx}
\usepackage{booktabs}
\usepackage{multirow}

\usepackage{empheq}

\usepackage{lipsum}

\usepackage[toc,page]{appendix}

\usepackage{rotating}

\usepackage{makecell}

\usepackage{lscape}
\usepackage{xcolor,colortbl}
\definecolor{lavender}{rgb}{0.9, 0.9, 0.98}
\everymath{\displaystyle}
\usepackage{float}
\usepackage[font={small}]{caption}
\usepackage[a4paper, total={6in, 8in}]{geometry}
\usepackage{csquotes}
\usepackage{url}
\newif\ifdouble

\newcommand{\papertitle}{Semi-Automated Generation and Hemodynamic Assessment of Surgical Baffle Geometry for Biventricular Repair}

\newcommand{\keywordOne}{computational cardiology}
\newcommand{\keywordTwo}{biventricular repair}
\newcommand{\keywordThree}{virtual surgery planning}
\newcommand{\keywordFour}{double outlet right ventricle}
\newcommand{\keywordFive}{patient-specific modeling}
\newcommand{\keywordSix}{computational fluid dynamics}
\doublefalse

\title{{\papertitle}}
\author{Elena Sabdy Martinez$^{1, 2}$, Alexander D. Kaiser$^{2,3,4}$, Alexander K. Reed$^{2,3}$, \\ Sascha W. Stocker$^{6}$, Amit Sharir$^{2,3}$, Perry S. Choi $^{2,3,4}$, Shiraz A. Maskatia$^{2,4,5}$, \\ Michael R. Ma$^{2,3,4}$, Alison Lesley Marsden$^{1, 2, 4, 5,6 *}$}

\date{\footnotesize
    $^1$ Institute for Computational and Mathematical Engineering, Stanford University, CA, USA \\
    $^2$ Cardiovascular Institute, Stanford University, CA, USA \\
    $^3$ Department of Cardiothoracic Surgery, School of Medicine, Stanford University, CA, USA \\
    $^4$ Maternal and Child Health Research Institute, Stanford University, CA, USA \\
    $^5$ Division of Pediatric Cardiology, Department of Pediatrics, School of Medicine, Stanford University, CA, USA \\
    $^6$ Department of Bioengineering, Stanford University, CA, USA \\
    $^*$ \textit{Corresponding author} (\texttt{amarsden@stanford.edu}) \\
}

\begin{document}
	\maketitle
	
	\begin{abstract}
		Patient-specific computational modeling has emerged as a powerful tool for surgical planning in complex congenital heart disease. One promising application is complex biventricular repair, which often requires construction of a custom intraventricular baffle to establish a physiologic left ventricle-to-aorta outflow pathway. In current practice, baffle geometry is designed and shaped intraoperatively. While existing preoperative modeling tools increasingly support visualization and virtual baffle design, the design process remains predominantly manual, and watertight models for quantitative hemodynamic assessment remain limited. We present a semi-automated computational framework for the design and assessment of patient-specific intraventricular baffles. The method constructs an explicit ventricular septal defect (VSD)-to-aorta flow pathway that preserves native right-ventricular myocardial and VSD geometry while reshaping the baffle region to satisfy patient-specific, section-wise area constraints. We retrospectively applied this framework to four patients with double outlet right ventricle (DORV) who had previously undergone biventricular repair. Each generated pathway was evaluated with pulsatile computational fluid dynamics (CFD), producing peak-flow pressure differences of 1.58--5.94 mmHg under the prescribed flow conditions. All modeled pressure differences remained below the 10-mmHg planning benchmark across mesh resolutions and under elevated flow. The current implementation preserves the native VSD and does not model VSD enlargement or conal resection; however, it provides a computational framework through which these interventions could be evaluated in silico. By translating a surgeon-defined baffle boundary into a physiologically constrained, simulation-ready model, this approach enables quantitative preoperative design and CFD-based evaluation of candidate baffle geometries, advancing patient-specific computational planning for biventricular flow restoration in heterogeneous anatomies.

	\end{abstract}
	
        \newpage
	\section{Introduction}
\label{sec:introduction}

     Congenital heart defects (CHDs) are the most common type of birth defect and remain among the leading causes of infant mortality worldwide \cite{sdf4chd, gbd2017}. CHDs encompass a broad and highly variable spectrum of structural abnormalities \cite{changing_landscape_chd}. Patients with complex defects may have anatomies in which only one ventricle effectively contributes to cardiac output, making it difficult to establish a two-ventricle circulation. Historically, many such patients were palliated to a single-ventricle physiology \cite{fontan_surgical_1971, hopkins_physiological_1985, norwood_fontan_1992, Kenny2018}. However, this strategy is associated with substantial long-term morbidity and mortality. Therefore, when anatomically feasible, several specialized centers offer complex biventricular repair as an alternative strategy that restores two-ventricle physiology \cite{rastelli_complete_1969, lecompte_reparation_1991, yacoub_anatomic_1984, Emani2012}. 

    In parallel, computational modeling has emerged as a powerful tool for generating patient-specific predictions across a range of clinical applications \cite{surrogate_modeling, Kadem2023, Niederer2019, Peirlinck2021, comp_model_chd_2} and supporting clinical decision making, with particular impact in pediatric cardiology \cite{taylor1999, kaiser2025, zuin2016, marsden2015, erdem2025, comp_model_chd, kaiser2024}. By improving understanding of complex cardiovascular flow patterns and integrating multimodal imaging data, these models can support patient-specific planning and comparison of surgical strategies \cite{marsden2015, erdem2025}. These advantages are particularly relevant in congenital anatomies requiring complex biventricular repair, where surgical reconstruction must be tailored to highly patient-specific morphology and substantial anatomical heterogeneity can make preoperative planning difficult.

    One particularly challenging defect is double outlet right ventricle (DORV), in which both the aorta and pulmonary artery arise predominantly from the right ventricle \cite{dorv_description}. DORV is also characterized by the presence of a ventricular septal defect (VSD), a hole between the ventricles that results in mixing of oxygenated and deoxygenated blood. Successful repair often requires the construction of an intraventricular tunnel that redirects left ventricular blood flow through the VSD to the aorta while avoiding flow obstruction \cite{dorv_4}. This is typically achieved using a surgically created intracardiac baffle, an internal patch or conduit that redirects blood flow within the heart \cite{dorv_1, dorv_2}. The VSD may be quite remote from the aorta, and the pathway may be crowded by cardiac structures, making the feasibility and design of the baffle highly anatomy-dependent \cite{dorv_3}. 
    
    Currently, baffles for DORV patients are designed and constructed intraoperatively on the arrested, unpressurized heart \cite{dorv_4}. Although 3D models and virtual planning tools are beginning to be used at some centers, baffle design remains largely dependent on surgeon experience and visual assessment, without a routine method for quantitatively evaluating predicted hemodynamic performance before surgery \cite{garekar2016, moore2018, slicerbaffle,priya2024,anilmaya2026}. The anatomical heterogeneity of DORV, variability in baffle design, and risk of postoperative obstruction therefore motivate the development of patient-specific computational planning tools.
    
    To address these challenges, a computational framework for patient-specific surgical planning requires (i) a three-dimensional (3D) anatomical model derived from imaging data, (ii) a method for generating a patch that satisfies defined anatomical and geometric constraints, and (iii) a method for evaluating its hemodynamic performance using computational fluid dynamics (CFD). Patient-specific cardiac models can now be routinely reconstructed from computed tomography (CT) or magnetic resonance imaging (MRI), with recent machine learning approaches enabling rapid whole-heart segmentation and reducing reliance on manual workflows \cite{huang2025, pandey2025, sdf4chd}. These models are used clinically for visualization, 3D printing, and surgical planning in complex congenital heart disease, including more recent virtual reality-based environments \cite{awori2021, erdem2025, slesnick2017, bhatla2016, hashemi2025}. Major advances toward preoperative planning have been made through interactive visualization tools that allow clinicians to inspect patient-specific anatomy and define candidate baffle pathways within reconstructed models \cite{slicerbaffle,priya2024,anilmaya2026}. In some cases, these geometries are subsequently flattened to support operative template preparation \cite{Kizilski2025, anilmaya2026}. Patient-specific DORV tunnels have also been manually constructed and evaluated prospectively using CFD \cite{capelli2018}, while idealized studies have examined the effects of tunnel geometry on pressure differences \cite{yang2020}. These studies establish the feasibility of virtual baffle planning, operative templates, and CFD evaluation, but they do not integrate a reproducible, constraint-based construction method with direct generation of a watertight domain for pulsatile CFD.

    This paper proposes a novel semi-automated computational pipeline for generating patient-specific intraventricular baffles that redirect flow from the left ventricle to the aorta. Using systolic-phase cardiac CT images, the method produces a 3D baffle that integrates directly with the patient’s anatomical model for CFD analysis and surgical planning. The resulting baffle forms a tunnel from the ventricular septal defect (VSD) to the aortic valve while maintaining sufficient cross-sectional area along its length.  To our knowledge, this is the first intraventricular baffle framework to combine reproducible, constraint-based geometry construction with generation of a simulation-ready CFD domain and pulsatile hemodynamic evaluation, extending baffle planning beyond visual assessment or geometric design alone. We retrospectively applied this framework to four DORV patients who had previously undergone biventricular repair at our institution and used pulsatile CFD to evaluate whether peak-flow pressure differences across the reconstructed pathways remained below a prespecified clinical planning benchmark. These results support the feasibility of the method and its potential use in preoperative planning.

	\section{Materials and Methods}
\label{sec:methods}
\begin{figure}[H]
\includegraphics[width=\textwidth]{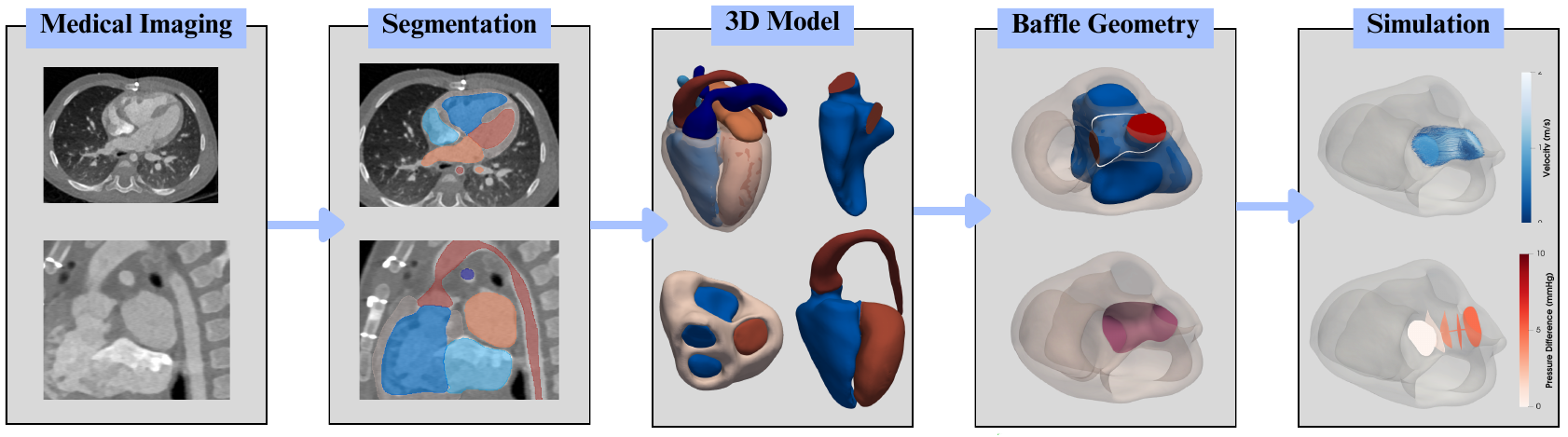}
   \caption{Computational workflow illustrated for patient 1. From left to right: \textbf{(a)} patient-specific cardiac imaging; \textbf{(b)} anatomical segmentation; \textbf{(c)} three-dimensional anatomical model and computational-domain preparation; \textbf{(d)} patient-specific baffle construction; and \textbf{(e)} pulsatile CFD evaluation. The baffle-integrated domain was prepared as a watertight, simulation-ready model.}
   \label{fig:overall_pipeline}
\end{figure}

We present a computational method to generate and evaluate patient-specific intracardiac baffles that establish biventricular circulation in congenital heart disease (CHD) patients. While this study focuses on  DORV with VSDs in varied anatomical locations, this methodology is broadly applicable to other complex CHDs involving intracardiac flow rerouting. For each patient, we performed the following steps: (1) reconstruction of a patient-specific 3D model from preoperative systolic CT imaging, (2) algorithmic generation of a closed intraventricular baffle surface, and (3) hemodynamic analysis of the reconstructed outflow pathway. An overview of this pipeline is illustrated in Figure~\ref{fig:overall_pipeline}.

\subsection{Patient cohort selection, preoperative imaging, and clinical data}
We retrospectively queried all patients from birth through 18 years of age with a clinical diagnosis of double outlet right ventricle (DORV) who underwent biventricular repair with an intracardiac left ventricular (LV) to aortic baffle at Lucile Packard Children’s Hospital from 2018--2025. Exclusion criteria included lack of preoperative computed tomography angiography (CTA) of sufficient quality for reconstruction, diagnosis of levo-transposition of the great arteries, and operative VSD enlargement. Cases involving VSD enlargement were excluded because the current framework preserves the native VSD geometry and does not model septal enlargement. A cohort of four patients was identified for preliminary feasibility evaluation (Table \ref{tab:cohort_selection}). All patients had postoperative transthoracic echocardiograms. Postoperative left ventricular outflow tract (LVOT) peak velocity, heart rate, and blood pressure were extracted from the final echocardiogram before discharge following biventricular repair.

\begin{table}
    \centering
    \small 
    \vspace{1mm}
\begin{tabularx}{\textwidth}{@{} p{0.3cm} p{0.35cm} p{0.3cm} p{3.5cm} X p{0.4cm} p{.7cm} p{.7cm} p{.7cm} p{.7cm} @{}}
        \toprule
        \textbf{Pt} & \textbf{Age (yrs)} & \textbf{Sex} & \textbf{Diagnoses} & \textbf{Surgical Procedure} & \textbf{Wt (kg)} & \textbf{CPB time (min)} & \textbf{XC time (min)} & \textbf{HR (bpm)} & \textbf{BP (mmHg)}\\ 
        \midrule
         1 & 1.13 & M & \{\textit{S},\textit{D},\textit{D}\} DORV, non-committed VSD, additional smaller inlet and apical VSDs, side-by-side outlet valves (Taussig-Bing) & Intracardiac DORV repair with LV-Ao baffle, additional VSD repair, patch augmentation of RVOT, sub-aortic conal and RVOT muscle resection, PA band removal with patch pulmonary arterioplasty & 9.4 & 339 & 242 & 131 & 101/70 \\
        \addlinespace
        2 & 0.82 & M & \{\textit{S},\textit{D},\textit{D}\} DORV, sub-aortic VSD, ASD, PDA, absent intrahepatic IVC, aortopulmonary collaterals to right lung & Intracardiac DORV repair with LV-Ao baffle, ASD closure, pulmonary valve commissurotomies, PA band removal with patch pulmonary arterioplasty, ligation of aortopulmonary collaterals, venoarterial ECMO cannulation & 7.35 & 231 & 158 & 133 & 107/78 \\
        \addlinespace
        3 & 0.44 & F & \{\textit{S},\textit{D},\textit{D}\} DORV with outlet VSD, pulmonary atresia, right aortic arch with vascular ring, bilateral SVC & Intracardiac DORV repair with LV-Ao baffle, RVOT muscle resection, RV-PA conduit insertion, central shunt ligation & 5.97 & 276 & 178 & 122 & 93/58 \\
        \addlinespace
        4 & 1.07 & F & \{\textit{S},\textit{D},\textit{D}\} DORV with sub-pulmonary VSD; additional apical VSD; doming, stenotic pulmonic valve; bilateral SVCs; ASD; PDA; single coronary artery & Intracardiac DORV repair with LV-Ao baffle, pulmonary root translocation with Gore-Tex interposition graft, pulmonary valve commissurotomies, conal septal muscle resection, ASD repair, PDA ligation, left SVC translocation & 7.8 & 243 & 164 & 136 & 87/53 \\
        \bottomrule
    \end{tabularx}
      \caption{DORV patient cohort data including postoperative heart rate and blood pressure. Segmental anatomy is depicted using the Van Praagh classification system in braces \cite{schallert2013}. \textit{Wt}: weight, \textit{CPB}: cardiopulmonary bypass, \textit{XC}: aortic cross-clamp, \textit{HR}: postoperative heart rate, \textit{bpm}: beats per minute, \textit{BP}: postoperative blood pressure, \textit{DORV}: double outlet right ventricle, \textit{VSD}: ventricular septal defect, \textit{ASD}: atrial septal defect, \textit{PDA}: patent ductus arteriosus, \textit{IVC}: inferior vena cava, \textit{d-TGA}: dextro-transposition of the great arteries, \textit{SVC}: superior vena cava, \textit{LV}: left ventricle, \textit{Ao}: aorta, \textit{RVOT}: right ventricular outflow tract, \textit{PA}: pulmonary artery, \textit{ECMO}: extracorporeal membrane oxygenation, \textit{RV-PA}: right ventricle-pulmonary artery}
    \label{tab:cohort_selection}
\end{table}

The Stanford University Institutional Review Board approved the study protocol and publication of data (IRB number: 39377, IRB approval date: June 17, 2025). Patient written consent for the publication of the study data was waived by the Institutional Review Board for use of anonymized, retrospectively acquired data.

\subsection{Computational domain preparation}
This section details the workflow used to transform medical imaging data into segmentations and subsequently into a three-dimensional, multi-labeled anatomical model and initial baffle. 

\subsubsection{Semi-automated machine learning segmentation} 
For each of the four patients, the end-systolic phase from the preoperative CTA was extracted and segmented into seven anatomic structures. The end-systolic phase was selected to capture ventricular geometry during peak ejection and better reflect the anatomy relevant to outflow reconstruction. The cardiac blood pools of the left ventricle (LV), right ventricle (RV), left atrium (LA), and right atrium (RA) were segmented to define the luminal portion of the heart, including trabeculations. The myocardial wall (myo) was segmented separately to distinguish tissue from blood pool. The pulmonary artery (PA) included the main pulmonary trunk and the proximal right and left branch PAs, while the aortic root (ao) consisted of the ascending aorta with the arch and branch vessels excluded.  

The images were initially segmented using a U-Net--based deep learning framework to reduce the manual effort required to generate each patient-specific anatomy. Convolutional neural networks are well established for supervised semantic segmentation, with U-Net--based models remaining highly competitive, especially in low-data regimes \cite{ronnebergerunet}. We employed the self-configuring nnU-Net framework \cite{isensee2021nnu}, which automates preprocessing, training configuration, and hyperparameter selection. The model was trained on the ImageCHD dataset (67 training, 5 validation, and 34 test cases) \cite{xu2021imagechd3dcomputedtomography, sdf4chd}. Training used the default nnU-Net settings with the DA5 data augmentation scheme, which applies extensive spatial and intensity transformations during training. This automated step was intended to accelerate initial model generation rather than replace clinical review.

The resulting segmentations, along with the corresponding CT images, were then imported into 3D Slicer for manual refinement. Mislabeled regions were corrected and unlabeled voxels were assigned to the appropriate anatomical class as needed, with particular attention to the ventricular septal interface and aortic valve annulus. All segmentations were reviewed by a cardiovascular imaging expert with subspecialty training in cardiovascular imaging for anatomic quality control and methodological consistency across patients.

\subsubsection{3D model generation and construction of a unified, capped flow domain}
The finalized multi-label segmentations were interpreted as three-dimensional image grids in which each voxel was assigned an integer label, ModelFaceID, corresponding to a specific anatomical structure. For each structure, a binary mask was generated by isolating voxels belonging to that label. To improve geometric and topological consistency, each binary mask underwent morphological closing and median filtering. A polygonal mesh, $\mathcal{M}_k$, was extracted using the marching cubes algorithm, post-processed to remove topological artifacts (e.g., non-manifold edges), and geometrically smoothed \cite{marching_cubes}. For each subdomain mesh $\mathcal{M}_k$, all triangular faces were assigned ModelFaceID = $k$ with the following mapping: LV = 1, RV = 2, LA = 3, RA = 4, myo = 5, ao = 6, PA = 7. 

For each patient, we constructed a single manifold mesh, $\mathcal{M}_{\text{AoBiV}}$, to represent the continuous blood pool of the LV, RV, and aorta by taking the union of the three segmented regions and following the same pipeline of filtering and surface extraction. This process automatically removed internal walls between constituent structures. We used the previously generated individual meshes $\{\mathcal{M}_k\}_{k\in\{1, 2, 6\}}$ as spatial references to assign anatomical labels to $\mathcal{M}_{\text{AoBiV}}$. ModelFaceID was assigned to each triangular face in $\mathcal{M}_{\text{AoBiV}}$ based on the nearest anatomical surface, determined from the distance between the face centroid and the corresponding labeled regions.

Explicit boundary surfaces corresponding to the aortic valve annulus and the VSD were generated by truncating $\mathcal{M}_{\text{AoBiV}}$ using intersection contours between labeled anatomical subsurfaces. Specifically, the LV-RV and RV-aorta interfaces were used to define the VSD and aortic valve cutting loops, respectively. Intersection curves were computed using a vtkIntersectionPolyDataFilter. Due to mesh discretization, these raw curves were often irregular and non-planar. To regularize them, radial rays were cast from a provisional centroid onto a best-fit plane, and the first ray-surface intersection in each direction was retained. The resulting ordered surface points were smoothed and reprojected onto the anatomical surface to form closed, non-self-intersecting cutting loops that remained coincident with the segmented anatomy.

\begin{figure}
\includegraphics[width=\textwidth]{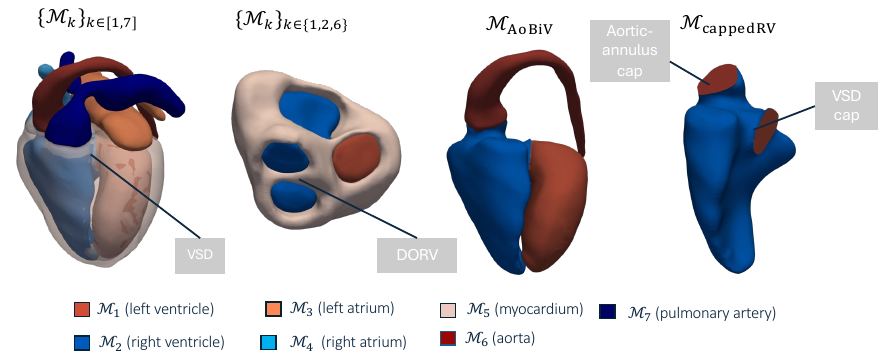
}
   \caption{Geometric preprocessing used to generate the computational domain, illustrated for patient 1. From left to right: \textbf{(a)} the initial patient-specific 3D anatomical model; \textbf{(b)} the separate LV, RV, and myocardial structures; \textbf{(c)} the unified LV--RV--aortic blood-pool model $\mathcal{M}_{\text{AoBiV}}$; and \textbf{(d)} the capped ventricular model $\mathcal{M}_{\text{cappedRV}}$, with planar caps at the VSD and aortic annulus.}
   \label{domain_pipeline}
\end{figure}

$\mathcal{M}_{\text{AoBiV}}$ was then truncated along each loop using a surface selection (vtkSelectPolyData) followed by a clipping operation (vtkClipPolyData), and the component bounded by both the VSD and aortic-annulus truncation loops was retained. The resulting VSD and aortic-annulus openings were capped with the SimVascular VMTK capping operation to produce a watertight domain comprising the native RV wall, a planar VSD cap, and a planar aortic-annulus cap. Each cap was assigned a unique ModelFaceID. The resulting multi-labeled domain, $\mathcal{M}_{\text{cappedRV}}$, formed the geometric basis for subsequent baffle construction. This preprocessing pipeline is summarized in Figure~\ref{domain_pipeline}.

\subsubsection{Definition of the baffle boundary and initial baffle surface} 
\label{sec:init_baff}
To determine the baffle attachment boundary, a congenital cardiac surgeon interactively defined the intended suture boundary for the given anatomy on the ventricular mesh using the BafflePlanner module in the SlicerHeart extension for 3D Slicer \cite{slicerheart, slicerbaffle}. The suture line was specified by placing $12-15$ discrete points in three-dimensional space, which delineated a clinically feasible border of the baffle along the endocardial surface of $\mathcal{M}_{\text{cappedRV}}$. The resulting coordinate set served as both a visual guide and as the computational input for the following automated contour generation.
The points were ordered in a principal-component plane, interpolated with a closed parametric spline, resampled at 600 locations, and projected onto $\mathcal{M}_{\text{cappedRV}}$ to obtain the smooth, surface-conforming attachment boundary $\mathcal{B}_{\text{baffle}}$. The ventricular surface was partitioned along this curve; the region containing both the VSD and aortic-annulus caps was retained and remeshed at a target edge length of 0.85 mm while preserving the VSD cap, aortic-annulus cap, ventricular wall, and attachment boundary as separate labels.

\begin{figure}
\includegraphics[width=\textwidth]{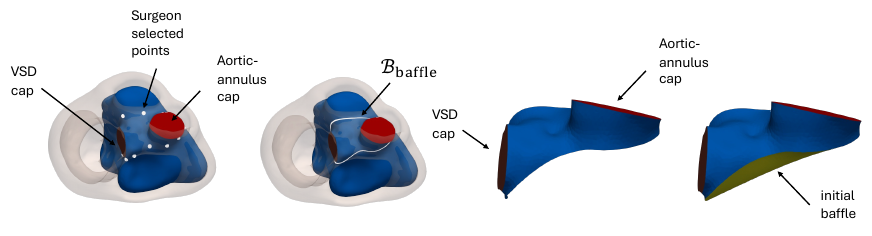}
   \caption{Initial-surface construction for patient 1. From left to right: \textbf{(a)} surgeon-selected baffle boundary points defining the intended patch attachment path on the ventricular geometry; \textbf{(b)} the complete, surface-conforming baffle boundary $\mathcal{B}_{\mathrm{baffle}}$; \textbf{(c)} the retained VSD-to-aortic ventricular surface after partitioning; and \textbf{(d)} the initial TPS baffle surface integrated with the retained ventricular surface to form the initial baffle-containing mesh.}
   \label{fig:initial_baffle}
\end{figure}

The opening was temporarily closed with a boundary-conforming thin-plate spline (TPS) patch (Figure~\ref{fig:initial_baffle}). The two sides of $\mathcal{B}_{\text{baffle}}$ were mapped by arc length to the perimeter of a structured unit disk, and nine equally spaced interior TPS control points were placed along the straight reference segment $\mathcal{R}_{\text{init}}$ joining boundary points adjacent to the VSD and aortic-annulus caps. TPS mapping produced the initial patch, after which its boundary vertices were restored exactly to $\mathcal{B}_{\text{baffle}}$. The complete construction was repeated using 3--29 interior control points. Across this range, the largest observed change in initial-surface position was 0.03 mm, and the corresponding downstream cross-sectional areas differed by less than 0.6\%. Nine control points were therefore used for all subsequent reconstructions. The joined surface was required to be watertight, connected, and free of non-manifold edges. This temporary surface provided a consistent parameterization and the reference segment used for centerline construction.

\subsection{Patient-specific baffle surface construction}
With the boundary fixed, the remaining task was to determine the extrusion depth of the baffle into the RV. Starting from the initial TPS surface, we constructed the final baffle surface, $\mathcal{S}_{\text{baffle}}$, and the integrated multi-surface domain, $\mathcal{M}_{\text{final}}$. The aim was to create a patient-specific intraventricular tunnel that achieved a prescribed target cross-sectional area, $A_{\text{target}}$, while maintaining a smooth, non-self-intersecting flow pathway.

\subsubsection{Determining $A_{\text{target}}$}
We aimed to produce a VSD-to-aorta tunnel that maintained a physiologically appropriate cross-sectional area throughout. We defined $A_{\text{target}}$ on a patient-specific basis to ensure that the resulting outflow geometry was consistent with normative values for body size. Clinical body surface area (BSA) values were provided by the clinical team and corresponded to the time of the index operation. The BSA-normalized reference annulus diameter at a z-score of 0 ($d_{z_0}$) was obtained from established pediatric cardiology nomograms. The corresponding cross-sectional area was computed assuming a circular orifice,
\begin{equation}
    A_{\text{target}}=\pi\left(d_{z_0}/2\right)^2.
\end{equation}

Normative values were derived using the Boston Children's Hospital BSA-aortic annulus z-score calculator \cite{colan2009normal, sluysmans2009structural}, which is derived from measurements of echocardiograms in normal children collected over approximately 12 years and follows established allometric adjustment methods. Anthropometric regressions (height, weight, BMI) use Centers for Disease Control and Prevention and World Health Organization reference datasets, and the echocardiographic z-score regressions are periodically updated as new normative models become available.  A z-score of 0 was selected because it represents the mean normal aortic-annulus caliber for body size and therefore provides a normative target for limiting postoperative LVOT obstruction. Patient-specific baffle target parameters for the four DORV cases are summarized in Table \ref{tab:patient_parameters}. 

\begin{table}[h!]
\centering
\caption{Patient-specific parameters for target area calculations.}
\label{tab:patient_parameters}
\begin{tabular}{cccc}
\toprule
Patient ID & BSA ($\text{m}^2$) & $d_{z_0}$ (cm)  & $A_{\text{target}}$($\text{cm}^2$)\\
\midrule
1 & 0.45 & 1.04 & 0.849 \\
2 & 0.37 & 0.94 & 0.694  \\
3 & 0.33 & 0.89 & 0.622 \\
4 & 0.39 & 0.97 & 0.739 \\
\bottomrule
\end{tabular}
\end{table}

\subsubsection{PCHIP-based centerline construction}
We next defined a centerline to provide a reproducible axis for cross-sectional construction. Medial-axis centerlines implemented in VMTK are commonly applied to tubular vascular lumens \cite{vmtk}, whereas the reconstructed intraventricular domain is broad and non-axisymmetric, with part of its boundary defined by the proposed baffle. Because the VSD and aortic-annulus caps and the surrounding ventricle are fixed in this domain, a medial-axis path may not approach the caps along their surface normals \cite{vtmk_centerline}. We therefore defined a custom construction axis with controlled endpoint tangents and stable section normals.

We constructed a smooth three-dimensional centerline connecting the VSD and aortic-annulus caps using a Piecewise Cubic Hermite Interpolating Polynomial (PCHIP) spline with normal-aligned endpoint transitions. The VSD and aortic-annulus cap centroids were denoted $\mathbf{p}_{\mathrm{in}}$ and $\mathbf{p}_{\mathrm{out}}$, respectively. Unit cap normals $\mathbf{n}_{\mathrm{in}}$ and $\mathbf{n}_{\mathrm{out}}$ were computed from the average surface normals and oriented into the domain, toward the opposite cap.

Near each cap, the path initially followed the inward-facing cap normal. For cap $c\in\{\mathrm{in},\mathrm{out}\}$, the nearest point $\mathbf{r}_c$ on the initial-surface reference segment $\mathcal{R}_{\mathrm{init}}$ defined in Section~\ref{sec:init_baff} was used to determine a geometry-based transition length $L_c$ and corresponding transition point $\mathbf{q}_c$:
\begin{equation}
    \mathbf{r}_c
    = \underset{\mathbf{x}\in\mathcal{R}_{\mathrm{init}}}{\operatorname{argmin}}\,
    \|\mathbf{x}-\mathbf{p}_c\|_2,
    \qquad
    L_c = (\mathbf{r}_c-\mathbf{p}_c)\cdot\mathbf{n}_c,
    \qquad
    \mathbf{q}_c=\mathbf{p}_c+L_c\mathbf{n}_c.
\end{equation}

This construction preserved cap-normal alignment and determined the transition directly from patient-specific geometry. To stabilize spline interpolation near the endpoints, two control points were specified outside the physical caps,
\begin{equation}
    \mathbf{p}_{0}
    = \mathbf{p}_{\mathrm{in}} - L_{\mathrm{in}}\mathbf{n}_{\mathrm{in}},
    \qquad
    \mathbf{p}_{N}
    = \mathbf{p}_{\mathrm{out}} - L_{\mathrm{out}}\mathbf{n}_{\mathrm{out}}.
\end{equation} The ordered control points
$\mathbf{P} = [\mathbf{p}_0,\mathbf{p}_{\mathrm{in}},\mathbf{q}_{\mathrm{in}},\mathbf{q}_{\mathrm{out}},\mathbf{p}_{\mathrm{out}},\mathbf{p}_N]$
were parameterized by cumulative chord length $s$, and three independent PCHIP interpolants, $x(s)$, $y(s)$, and $z(s)$, were fit to obtain the parametric centerline
\begin{equation}
    \mathbf{c}(s) = [x(s), y(s), z(s)]^T.
\end{equation}

PCHIP was selected because its shape-preserving cubic interpolation provides a continuous first derivative while limiting overshoot between the prescribed control points \cite{pchip}. The physical centerline was restricted to $[\mathbf{p}_{\mathrm{in}},\mathbf{p}_{\mathrm{out}}]$ and sampled at an arc-length spacing of 0.50 mm. Analytic PCHIP derivatives defined the section normals independently of sampling density. To test sampling-density sensitivity, the complete geometry construction was repeated at 0.25-, 0.50-, and 1.00-mm spacing while holding the attachment boundary, endpoint-transition rule, and target area fixed. All 12 reconstructions retained the target-area criterion. Relative to the 0.50-mm reconstruction, the maximum centerline displacement was 0.064 mm and total baffle surface area differed by at most 0.12\%. Figure~\ref{fig:centerline}(a) shows the ordered control points and initial-surface reference segment, and Figure~\ref{fig:centerline}(b) shows the resulting PCHIP construction axis.

\subsubsection{Cross-sectional contour extraction}
\begin{figure}[H]
\includegraphics[width=\textwidth]{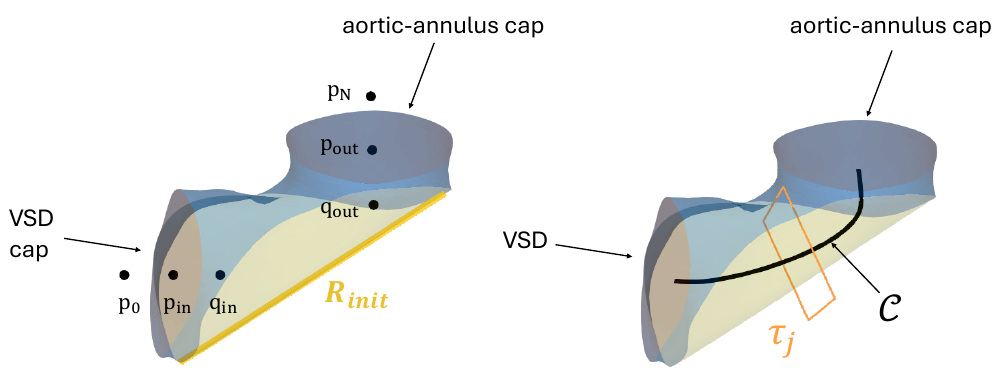}
   \caption{Custom centerline construction for patient 1. \textbf{(a)} Ordered control points $\mathbf{P}$ and the initial-surface reference segment $\mathcal{R}_{\mathrm{init}}$; and \textbf{(b)} the resulting PCHIP centerline $\mathcal{C}$ and a representative section plane $\Pi_j$ orthogonal to its analytic tangent $\mathbf{t}_j$. Points $\mathbf{p}_0$ and $\mathbf{p}_N$ lie beyond the VSD and aortic-annulus caps to stabilize the endpoint directions; only the portion from $\mathbf{p}_{\mathrm{in}}$ to $\mathbf{p}_{\mathrm{out}}$ is sampled.}
      \label{fig:centerline}
\end{figure}

For each centerline point $\mathbf{p}_j=\mathbf{c}(s_j)\in\mathcal{C}$, the initial capped domain was intersected with a plane orthogonal to the local centerline direction. The unit tangent was calculated from the analytic PCHIP derivative:
\begin{equation}
    \mathbf{t}_j=\frac{\mathbf{c}'(s_j)}{\|\mathbf{c}'(s_j)\|_2}.
\end{equation}
The corresponding cutting plane was defined as
\begin{equation}
    \Pi_j = \{ \mathbf{x} \in \mathbb{R}^3 \mid \mathbf{t}_j \cdot (\mathbf{x} - \mathbf{p}_j) = 0 \}.
\end{equation}
A representative centerline-orthogonal section plane is shown in Figure~\ref{fig:centerline}(b).

The plane--surface intersection produced one or more closed contours. When multiple contours were present, the contour whose centroid was nearest $\mathbf{p}_j$ was retained as $\mathcal{L}_j$, and each of its vertices inherited the ModelFaceID of its nearest surface triangle. Contours with more than 75$\%$ of their vertices on the native RV wall were excluded because they did not meaningfully sample the baffle. Each retained contour was split at the two locations nearest $\mathcal{B}_{\text{baffle}}$. The arc containing the majority of baffle-labeled vertices was classified as the baffle segment $\mathcal{A}_{j, \text{baffle}}$, and the complementary arc was classified as the fixed RV segment $\mathcal{A}_{j, \text{fixed}}$. 

\subsubsection{Cross-sectional shape adjustment to achieve target area} 
The baffle segment $\mathcal{A}_{j, \text{baffle}}$ was then adjusted, while $\mathcal{A}_{j, \text{fixed}}$ was held constant. The objective was to reshape $\mathcal{A}_{j, \text{baffle}}$ so that, when combined with $\mathcal{A}_{j, \text{fixed}}$, the resulting closed loop enclosed an area no smaller than the prescribed $A_{\text{target}}$. The segment $\mathcal{A}_{j, \text{baffle}}$ was first projected onto its best-fit plane, obtained via singular value decomposition of the loop vertices. All subsequent operations were performed in this planar domain.  Let $\mathbf{q}_{0j}(s)$ denote the projected original baffle segment, parameterized by normalized cumulative arc length $s\in[0,1]$. The segment was parameterized by a scalar amplitude $\alpha_j \geq 0$ using a parabolic displacement normal to the chord connecting its endpoints: 
\begin{equation}
    \mathbf{q}_j(s;\alpha_j)
=
\mathbf{q}_{0j}(s)
+
\alpha_j\,4s(1-s)\mathbf{v}_j,
\qquad s\in[0,1],
\end{equation}
where $\mathbf{v}_j$ was a unit normal to the endpoint chord, within the fitted plane and oriented toward increasing enclosed area. 

The updated loop $\mathcal{L}_j(\alpha_j)$ was formed by combining the unchanged fixed segment $\mathcal{A}_{j, \text{fixed}}$ with the modified baffle segment $\mathcal{A}_{j, \text{baffle}}(\alpha_j)$. A 0.5\% construction margin was applied to account for small geometric changes introduced during final-surface reconstruction. The amplitude was determined by solving the one-dimensional constrained problem
\begin{equation}
\begin{aligned}
    \underset{\alpha_j \geq 0}{\operatorname{minimize}} \quad & \alpha_j \\
    \text{subject to} \quad & \operatorname{Area}\!\left(\mathcal{L}_j(\alpha_j)\right) \geq 1.005A_{\text{target}}.
\end{aligned}
\end{equation}
If the unmodified section satisfied the construction target, $\alpha_j$ was set to zero. Otherwise, because the prescribed displacement produced a linear change in enclosed area with $\alpha_j$, the minimum feasible amplitude was computed directly for each cross-section.

\subsubsection{One-sided envelope regularization}
The raw amplitudes $\{\alpha_j\}$ obtained from the cross-sectional optimization ensured that each reconstructed loop individually satisfied the target area constraint. However, directly applying these amplitudes produced high-frequency variations along the centerline, leading to local inward curvature between adjacent sections and an irregular baffle surface. These geometric irregularities created abrupt changes in cross-sectional area and surface normal direction, which are undesirable for smooth ventricular outflow and inconsistent with the smooth configurations expected of pericardial baffles \cite{baffle_material, Aguiari2016}. 

To mitigate these geometric irregularities, we imposed a concavity constraint on the amplitude sequence. Specifically, we computed the least concave majorant in centerline arc length: the smallest piecewise-linear function with non-increasing slope that lay at or above all required amplitudes. Let $i_1 < i_2 < \cdots < i_K$ denote its vertices. The regularized amplitudes were

\begin{equation} 
    \hat{\alpha}_j = \alpha_{i_k} + \frac{\alpha_{i_{k+1}}- \alpha_{i_k}}{s_{i_{k+1}}-s_{i_k}}\,(s_j-s_{i_k}), \qquad i_k \le j \le i_{k+1}.
\end{equation}

The one-sided property, $\hat{\alpha}_j\geq\alpha_j$, ensured that regularization did not decrease the section-wise displacement required to satisfy the area constraint, and the target remained satisfied at every section. Geometrically, this enforced a smooth, gradually varying tunnel profile, avoiding abrupt changes in cross-sectional area and curvature.

\subsubsection{Baffle surface generation via thin-plate spline interpolation}

The final baffle surface was generated by adapting a landmark-based Thin-Plate Spline (TPS) spatial mapping $f: \mathbb{R}^2 \rightarrow \mathbb{R}^3$ \cite{tps}, a technique commonly used for smooth landmark-based interpolation and shape deformation \cite{slicerbaffle}. The two-dimensional domain was the structured unit-disk parameterization established during construction of the initial baffle surface, so that the reconstruction inherited its connectivity and boundary correspondence. We defined two sets of three-dimensional target landmarks for the TPS interpolation: (i) the baffle attachment boundary $\mathcal{B}_{\text{baffle}}$ and (ii) the ordered set of final cross-sectional loops representing the desired baffle profile $\{\hat{\mathcal{L}}_j\}_{j=1}^M$. The TPS mapping interpolated these landmarks while minimizing a bending energy functional \cite{slicerheart, tps_use, tps_use_2}. Boundary landmarks were restored to their prescribed coordinates so the attachment seam was reproduced exactly. The resulting surface $\mathcal{S}_{\text{baffle}}$ was subsequently represented by its piecewise-linear triangulation and assigned a unique ModelFaceID.

\begin{figure}[H]
\includegraphics[width=\textwidth]{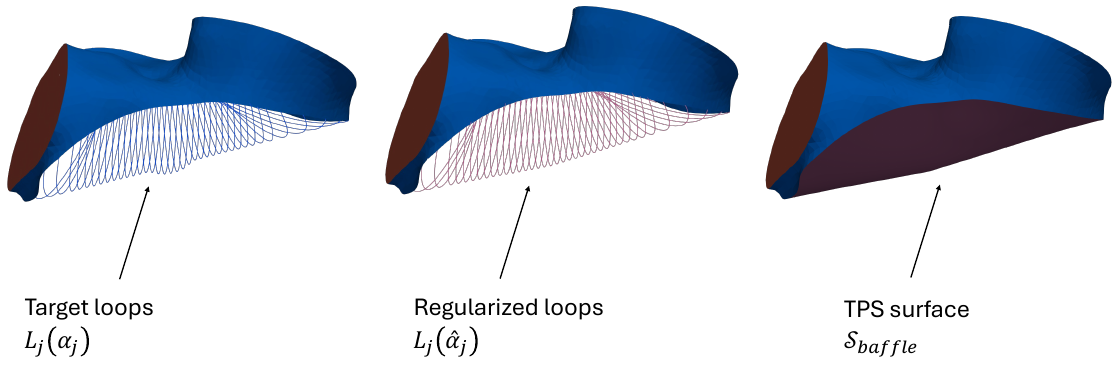}
   \caption{Section constraints and final surface for patient 1. From left to right: \textbf{(a)} cross-sectional loops $\mathcal{L}_j(\alpha_j)$ satisfying the prescribed area target; \textbf{(b)} one-sided envelope-regularized loops $\mathcal{L}_j(\hat{\alpha}_j)$; and \textbf{(c)} the final TPS baffle surface $\mathcal{S}_{\text{baffle}}$, reconstructed from the regularized profiles while preserving the surgeon-defined attachment boundary.}
\label{fig:loops}
\end{figure}

This surface was merged with the reconstructed ventricular mesh along $\mathcal{B}_{\text{baffle}}$, replacing the initial baffle and yielding a watertight surface mesh $\mathcal{M}_{\text{final}}$.  Post-processing in SimVascular included mesh cleaning, computation of consistent surface normals, and isotropic remeshing to obtain a high-quality, simulation-ready model. The final mesh consisted of the VSD cap, aortic-annulus cap, native right-ventricular wall, and the constructed baffle surface, with each region uniquely identified by its ModelFaceID. The baffle surface, $\mathcal{S}_{\text{baffle}}$, was extracted from $\mathcal{M}_{\text{final}}$ using this labeling while remaining an integral component of the full mesh. A congenital cardiac surgeon reviewed each final mesh to confirm that its baffle geometry was within the range of surgically feasible configurations under current practice, providing an additional clinical feasibility check. Figure~\ref{fig:loops} summarizes the progression from area-constrained section loops to the regularized loops and final TPS surface.

\subsection{Hemodynamic simulation}
The final baffle-integrated domain $\mathcal{M}_{\text{final}}$ was a closed, simulation-ready representation of the reconstructed VSD-to-aortic pathway. For the simulations reported here, $\mathcal{M}_{\text{final}}$ was extended from a plane through the mid-left-ventricular cavity to the sinotubular junction. These reproducible anatomical landmarks placed the extended-domain inflow and outflow boundaries away from the reconstructed pathway and its measurement planes, reducing the direct influence of the prescribed boundary conditions on the local assessment. The area-constrained portion of the baffle remained unchanged, while peripheral attachment remeshing changed total patch area by less than 0.03\%. The VSD and aortic-annulus caps used for geometric construction were retained as internal measurement planes, so the reported pressure difference characterized the reconstructed pathway rather than the added extensions.

To evaluate the hemodynamics of our final designs, we simulated three-dimensional, pulsatile blood flow within the extended domain. All simulations used the same numerical and fluid-property parameters. Blood flow was modeled as an incompressible, Newtonian fluid governed by the Navier--Stokes equations:
\begin{align}
\rho \left( \frac{\partial \mathbf{u}}{\partial t} + (\mathbf{u} \cdot \nabla)\mathbf{u} \right) &= -\nabla p + \mu \nabla^2 \mathbf{u} \\
\nabla \cdot \mathbf{u} &= 0 & \text{in } \Omega,
\end{align}

where $\Omega$ is the extended fluid domain, $\mathbf{u}$ is the fluid velocity vector, and $p$ is the pressure. The fluid density was set to $\rho = 1.06~\mathrm{g/cm^3}$ and the dynamic viscosity to $\mu = 0.04~\mathrm{g/(cm\cdot s)}$ \cite{morris2015}.

To define a consistent, body-size-scaled inflow condition across patients, cardiac output was calculated as \begin{equation}
\mathrm{CO}=\mathrm{CI}_{\mathrm{ref}}\times\mathrm{BSA}
\end{equation} using a normative cardiac index, $\mathrm{CI}_{\mathrm{ref}}=3.48~\mathrm{L\,min^{-1}\,m^{-2}}$, representing the reported mean for healthy infants followed through the first year of life \cite{alverson}. The corresponding mean volumetric flow rate was calculated as $Q_{\mathrm{mean}}=1000\,\mathrm{CO}/60$ (cm$^3$/s). This provided a reproducible, body-size-scaled condition for evaluating whether each reconstructed pathway could accommodate the normative systemic flow that the repair was intended to restore.

The resulting mean flow was used to scale the pulsatile inflow condition. The flow profile was adapted from a deidentified healthy reference waveform acquired at our institution with an inherent period of 0.682 s, equivalent to approximately 88 beats/min. The waveform was normalized by its cycle integral and scaled for each patient so that the cycle volume equaled $Q_{\mathrm{mean}}T$, where $T=0.682$ s.

The computational domain boundary was divided into three regions: the extended-domain inflow boundary, the extended-domain outflow boundary, and the enclosing rigid walls, comprising the native ventricular wall and reconstructed baffle surface. Wall surfaces were modeled as rigid. A spatially uniform (plug) velocity profile was prescribed at the extended-domain inflow boundary because the truncated ventricular cavity does not represent a long, straight tube in which a fully developed parabolic profile would be expected. A zero-velocity Dirichlet condition was applied to the rigid walls. At the extended-domain outflow boundary, a resistance boundary condition was imposed to represent the downstream vasculature, with $R =\mathrm{MAP}/Q_{\text{mean}}$, so that the cycle-averaged outlet pressure matched the prescribed patient-specific mean arterial pressure. Mean arterial pressure was estimated from the postoperative systolic and diastolic blood pressures as $\mathrm{MAP}=(\mathrm{SBP}+2\mathrm{DBP})/3$. With prescribed inflow and a single resistance outlet, MAP set the mean outlet-pressure level and was not used to prescribe flow or as the primary hemodynamic outcome. A nominal or clinically selected MAP can therefore be specified when the workflow is applied prospectively. The resulting inflow rates and outlet resistances are summarized in Table~\ref{tab:hemodynamics}.

\begin{table}[h!]
\centering
\caption{Normative flow and patient-specific pressure inputs. }
\renewcommand{\arraystretch}{1.2}
\begin{tabular}{ccccccc}
\toprule
\textbf{Patient} & \textbf{CO} & \textbf{$Q_{\text{mean}}$} & \textbf{$Q_{\text{peak}}$} &\textbf{MAP} &  \textbf{Resistance}  &  \textbf{Resistance}  \\
 & \textbf{(L/min)} & \textbf{(cm$^3$/s)} &  \textbf{(cm$^3$/s)} &  \textbf{(mmHg)} & \textbf{(dyne$\cdot$s/cm$^5$)} & \textbf{(Wood)}\\
\midrule
1 & 1.57 & 26.1 & 85.2 & 80 & 4087 & 51.1\\
2 & 1.29 & 21.5 & 70.0 & 88 & 5467 & 68.3 \\
3 & 1.15 & 19.1 & 62.5 & 70 & 4876 & 61.0\\
4 & 1.36 & 22.6 & 73.8 & 64 & 3772 & 47.1 \\
\bottomrule
\end{tabular}
\label{tab:hemodynamics}
\end{table}

Simulations were performed using svMultiPhysics, a high-performance finite-element solver within the SimVascular software suite \cite{svmultiphys}.  The coupled velocity-pressure system was discretized using a stabilized finite-element formulation \cite{esmaily_bipartition} and solved with a preconditioned GMRES iterative solver, with FSILS used as the preconditioner and a relative tolerance of $10^{-4}$, a maximum of 150 linear-solver iterations, and a backflow stabilization coefficient of 0.2. Each pulsatile simulation consisted of four complete 0.682-s cycles with a time step of 0.001 s. Spatial sensitivity was evaluated using coarse, nominal, and fine target edge lengths of 1.25, 0.95, and 0.75 mm. Nominal volume-mesh quality was assessed using the minimum and 1st-percentile scaled Jacobian and the 99th-percentile aspect ratio for each patient. Positive scaled Jacobians confirmed that no tetrahedra were inverted, while aspect ratio characterized element elongation. To test higher-flow sensitivity, each patient was also simulated on the nominal mesh at a cardiac index of $4.22~\mathrm{L\,min^{-1}\,m^{-2}}$, corresponding to the normative mean plus one standard deviation. The complete pulsatile waveform was rescaled to this cardiac index, and outlet resistance was recalculated to preserve the prescribed mean arterial pressure \cite{esmaily_resistance}.

Hemodynamic outcomes were evaluated at the saved fourth-cycle time point corresponding to peak prescribed inflow ($t=2.210$ s). The pressure difference was defined as the face-averaged pressure on the VSD measurement plane minus that on the aortic-annulus measurement plane. Separately from the PCHIP construction axis, a medial flow-domain centerline was generated from the extended CFD lumen using the SimVascular implementation of VMTK and trimmed to the interval between the VSD and aortic-annulus measurement planes \cite{vmtk,vtmk_centerline}. Analyzed pathway length was defined as the arc length of this centerline, and maximum centerline velocity was the greatest pointwise velocity sampled along it. A prespecified, clinician-defined pressure-difference threshold of 10 mmHg was used as a planning criterion, consistent with congenital literature classifying Doppler peak LVOT gradients $\leq 10$ mmHg as normal \cite{vanderpalen2025}. Postoperative Doppler LVOT peak velocity was reported for relative comparison.

	\section{Results}
The semi-automated baffle generation and simulation workflow was applied retrospectively to the preoperative imaging data from four DORV patients who had previously undergone biventricular repair with an intracardiac LV--Ao baffle. For each case, the reconstructed flow pathway $\mathcal{M}_{\text{final}}$, including the generated baffle surface $\mathcal{S}_{\text{baffle}}$, was extended and evaluated using pulsatile CFD. All finalized patient-specific surface models and SimVascular-compatible tetrahedral volume meshes for all four patients are publicly available at https://github.com/StanfordCBCL/Surgical-Baffle-Geometries. 

\subsection{Final baffle geometries and model quality}

\begin{figure}
\includegraphics[width=\textwidth]{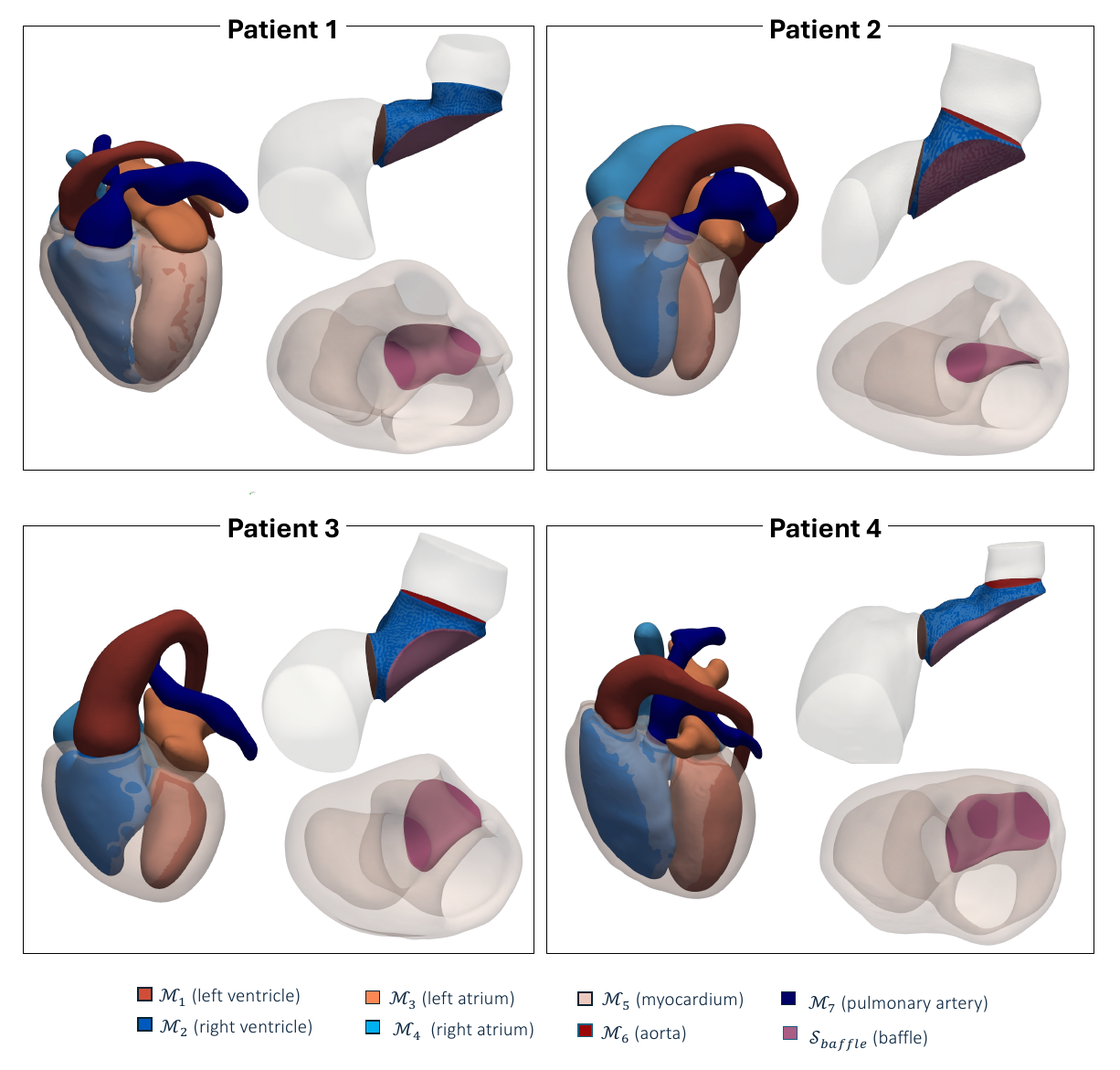}
\caption{Patient-specific reconstructions for the four DORV patients. Within each patient panel, the left image shows the reconstructed cardiac anatomy; the upper-right image shows the reconstructed VSD-to-aortic pathway $\mathcal{M}_{\text{final}}$ with the simulation extensions shown translucently; and the lower-right image shows the baffle surface $\mathcal{S}_{\text{baffle}}$ relative to the surrounding myocardium.} \label{fig:pipeline_for_all_patients}
\end{figure}

The workflow generated patient-specific surface models and simulation-ready meshes for all four cases. Each baffle surface was continuous, free of self-intersections, and coincident with the anatomy along the attachment boundary. All four integrated domains were watertight, free of non-manifold edges, and discretized with tetrahedral volume meshes. Figure~\ref{fig:pipeline_for_all_patients} shows the reconstructed anatomy, the reconstructed pathway within the extended simulation domain, and the anatomical position of the baffle for each patient. Nominal volume-mesh quality was assessed using scaled Jacobian and aspect ratio. All tetrahedra had positive scaled Jacobians, confirming that none were inverted; patient-specific mesh-quality values are reported in Table~\ref{tab:validation}.

\begin{table}
\centering
\caption{Geometric targets, model dimensions, and nominal volume-mesh quality for the final reconstructions.}
\label{tab:validation}
\renewcommand{\arraystretch}{1.15}
{
\begin{tabular}{lcccc}
\toprule
& \textbf{Patient 1}
& \textbf{Patient 2}
& \textbf{Patient 3}
& \textbf{Patient 4} \\
\midrule
Target area $A_{\text{target}}$ (mm$^2$)
& 84.95 & 69.40 & 62.21 & 73.90 \\
Flow-domain centerline length (mm)
& 19.16 & 10.14 & 13.31 & 26.15 \\
Baffle patch area (mm$^2$)
& 564.8 & 255.5 & 403.1 & 763.5 \\
Baffle surface triangles
& 3596 & 1955 & 2425 & 5460 \\
Nominal volume-mesh tetrahedra ($10^3$)
& 333 & 101 & 150 & 457 \\
Minimum scaled Jacobian
& 0.101 & 0.109 & 0.097 & 0.084 \\
1st-percentile scaled Jacobian
& 0.215 & 0.213 & 0.216 & 0.215 \\
99th-percentile aspect ratio
& 4.41 & 4.41 & 4.42 & 4.43 \\
\bottomrule
\end{tabular}
}
\end{table}

\subsection{{Pulsatile hemodynamic assessment}}

Pulsatile flow was successfully simulated in the extended domain for all four reconstructed pathways. The primary analysis used the nominal 0.95-mm mesh and the saved fourth-cycle time point corresponding to peak prescribed inflow ($t=2.210$ s). Across all simulations, cycle-3 and cycle-4 pressure-difference waveforms at the extended-domain inflow and outflow boundaries had a normalized root-mean-square difference of at most 0.231\%, supporting periodicity by the fourth cycle.

\begin{figure}
\includegraphics[width=\textwidth,trim=30 45 30 45,clip]{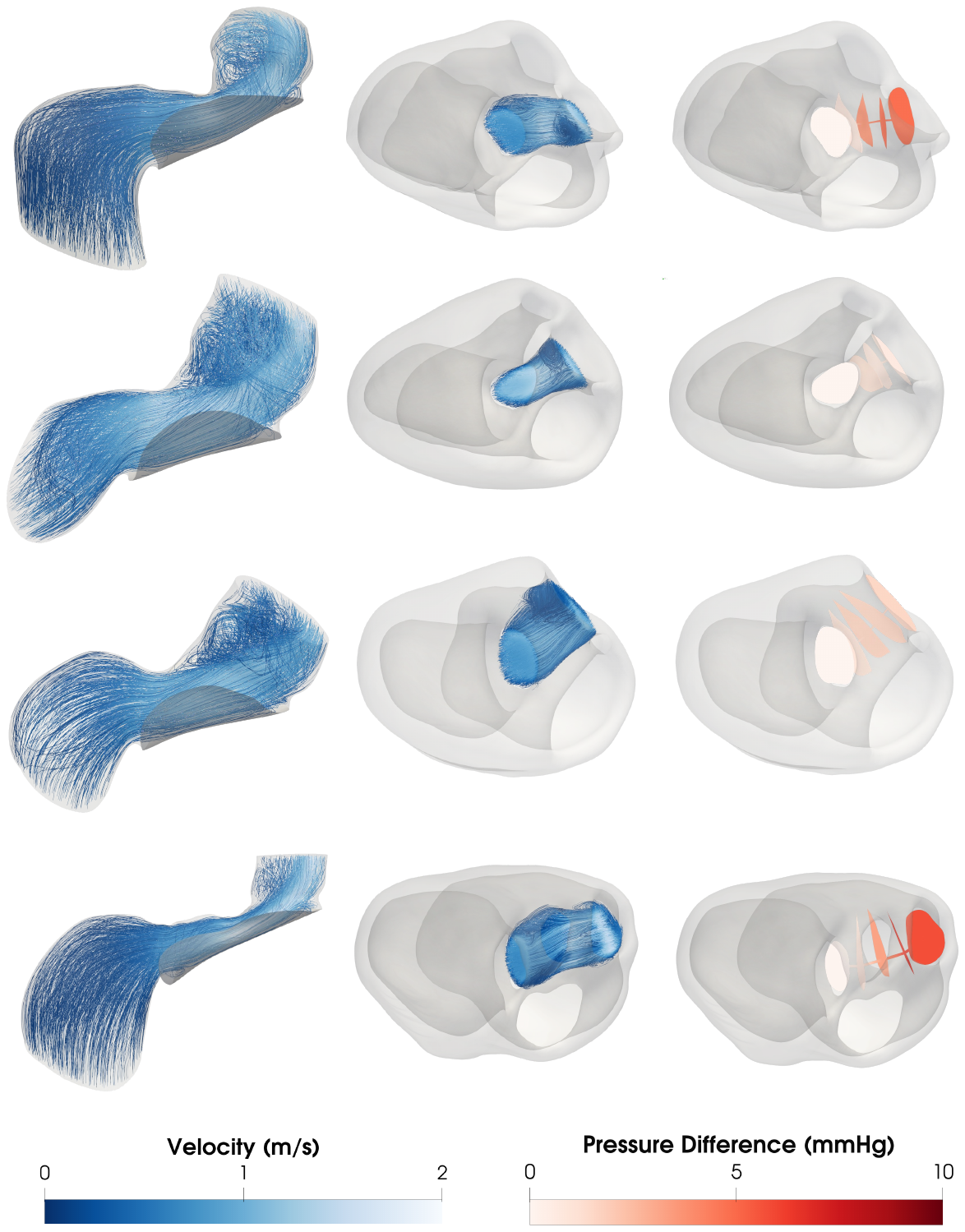}
    \caption{Peak-flow visualization for the four patient-specific models (patients 1--4 by row). From left to right: streamlines through the extended simulation domain, colored by velocity magnitude; velocity magnitude through the reconstructed pathway within the surrounding ventricular anatomy; and pressure difference relative to the VSD shown on five cross-sectional planes. The pressure planes include the VSD and aortic-annulus measurement planes and three equally spaced internal planes along the flow-domain centerline. Common color limits are shown at the bottom.}
   \label{fig:sims}
\end{figure}

The face-averaged pressure difference was the primary functional outcome because it quantified the pressure change across the reconstructed pathway. At the analyzed peak-flow time point under the nominal cardiac-index condition, pressure differences ranged from 1.58 to 5.94 mmHg and remained below the 10-mmHg planning benchmark in all four patients. Maximum centerline velocity was evaluated as a consistent measure of flow acceleration. Modeled maximum centerline velocities ranged from 1.14 to 1.62 m/s. For comparison, postoperative LVOT peak velocities measured by continuous-wave Doppler echocardiography ranged from 1.11 to 1.78 m/s (Table~\ref{tab:sim_vals}). Continuous-wave Doppler records the maximum flow velocity detected along the entire beam path at the expense of spatial resolution, preventing precise localization while preserving velocity data from aliasing \cite{quinones2002}. The CFD metric represents the maximum pointwise velocity sampled along the flow-domain medial centerline. Although these metrics are not directly equivalent, their similar order of magnitude supports the modeled velocity scale. The combined pressure and velocity results indicate that the reconstructed pathways accommodated the prescribed systemic flow within the selected planning criteria. Figure~\ref{fig:sims} shows the peak-flow velocity and pressure-difference fields for the nominal meshes.

\begin{table}[H]
\centering
\caption{Peak-flow pressure differences and velocity comparison.}
\label{tab:sim_vals}
{
\small
\begin{tabular}{cccc}
\toprule
Patient & CFD $\Delta P$ (mmHg) & CFD centerline $v_{\max}$ (m/s) & Echo LVOT $v_{\max}$ (m/s) \\
\midrule
1 & 4.81 & 1.43 & 1.78 \\
2 & 1.58 & 1.48 & 1.47 \\
3 & 1.68 & 1.14 & 1.40 \\
4 & 5.94 & 1.62 & 1.11 \\
\bottomrule
\end{tabular}
}
\end{table}

Complete coarse, nominal, and fine comparisons were available for all four patients. Nominal-minus-fine differences ranged from $-0.18$ to $+0.29$ mmHg for peak-flow pressure difference and from $+0.006$ to $+0.097$ m/s for maximum centerline velocity. All pressure differences at every mesh level remained below the 10-mmHg planning benchmark. Thus, spatial discretization did not change the primary pressure-difference conclusion. The conclusion was also unchanged under the elevated-flow condition at a cardiac index of 4.22~$\mathrm{L\,min^{-1}\,m^{-2}}$, for which peak-flow pressure differences ranged from 2.37 to 8.85 mmHg and maximum centerline velocities ranged from 1.38 to 1.96 m/s.

	\section{Discussion}
\label{sec:discussion}
Computational modeling has emerged as an essential tool for patient-specific analysis of congenital heart defects, with applications spanning electrophysiology, electromechanics, solid mechanics, and fluid dynamics \cite{gray2018, clayton2008, reumann2009,salvador2024,taylor2009,shad2022}. Baffle geometry for complex biventricular repair, however, is still largely determined intraoperatively on an arrested, flaccid heart, and the surgeon must estimate how the outflow baffle will perform once filled with blood, pressurized, and reanimated after repair. Here, we describe a robust workflow to generate clinically reasonable baffle geometries that are integrated with CFD analysis. This methodology preserves the surgeon-specified baffle boundary, modifies only the proposed baffle segment to meet a patient-size-specific cross-sectional area target, and directly produces a watertight domain for CFD analysis. To our knowledge, this is the first intraventricular baffle framework to integrate constraint-based geometry construction with pulsatile CFD evaluation. In four DORV patients, the generated pathways satisfied the prescribed area constraint, were considered clinically plausible by a congenital cardiac surgeon, and maintained peak-flow pressure differences below the planning benchmark across mesh resolutions and under elevated flow. These findings show that baffle planning can extend beyond visual assessment to a reproducible, anatomically grounded design that can be quantitatively evaluated before surgery.

This work builds on prior approaches to preoperative planning for complex congenital heart disease. Three-dimensional reconstructions and printed models improve visualization of the VSD, great arteries, and potential tunnel trajectory \cite{garekar2016, moore2018, liang2022}. Interactive and virtual-reality methods allow candidate baffles to be created, revised, and measured \cite{slicerbaffle, priya2024}, and recent work has translated virtual designs into operative templates \cite{slicerbaffle, anilmaya2026}. These approaches establish the feasibility of patient-specific baffle planning but do not provide a reproducible section-wise area constraint or a fully surface-conforming attachment boundary. An important innovation of this workflow is that the user-defined baffle boundary is constrained to remain coincident with the underlying anatomical mesh, enabling robust generation of a closed, watertight geometry suitable for CFD analysis. In the current implementation, the boundary was defined interactively by a congenital cardiac surgeon, who selects control points along the intended patch suture line in a 3D rendering. These control points were converted into a complete, surface-conforming suture line while preserving the surgeon-selected path.

The VSD-to-aorta flow domain is a broad, non-tubular ventricular cavity, so centerline methods developed for tubular vascular domains may not provide the desired cap-normal directions at the inlet and outlet \cite{vmtk,vtmk_centerline}. We therefore developed a custom PCHIP-based construction in which the inlet and outlet tangent directions were prescribed and aligned with the inward cap normals, while analytic PCHIP derivatives defined consistent section normals \cite{pchip}. At each cross-section, only the surgically created baffle segment was modified to meet the target area, preserving the native right-ventricular myocardial segment of the reconstructed pathway. Because independently enforcing the target area can create abrupt section-to-section changes, a one-sided concave envelope was used to regularize the longitudinal profile while maintaining the required area at each section. This geometric step was intended to avoid sharp geometric transitions that could promote local flow separation or recirculation. The final baffle surface was constructed using a thin-plate spline (TPS) formulation that interpolated the prescribed boundary curve and the regularized cross-sectional profiles. TPS was selected because it produces a minimum-bending-energy surface while satisfying the prescribed boundary constraints \cite{tps, slicerbaffle}. The resulting construction translates section-wise area requirements into a continuous, boundary-conforming baffle surface that retains explicit control of endpoint geometry and is suitable for surgeon review and CFD meshing.

Pulsatile CFD moves baffle planning beyond visual assessment by enabling quantitative evaluation of the reconstructed pathway. Prior work demonstrated the feasibility of using CFD to evaluate manually constructed, patient-specific DORV tunnels \cite{capelli2018}. The present method directly couples reproducible, area-constrained construction with a face-averaged VSD-to-aortic pressure-difference metric. At nominal flow, peak-flow pressure differences were 1.58--5.94 mmHg and remained below the prespecified 10-mmHg clinical planning benchmark in all four patients. The coarse, nominal, and fine volume-mesh comparison assessed spatial-discretization sensitivity, while the elevated-flow condition at a cardiac index of 4.22 L/min/m$^2$ tested whether a plausible higher-flow state changed the pressure-difference conclusion; all modeled peak-flow pressure differences remained below the planning benchmark. Clinically, residual LVOT obstruction is assessed using peak Doppler gradients and, when available, catheter-derived pressure differences. Congenital series have classified Doppler peak LVOT gradients $\leq 10$ mmHg as normal and used postoperative gradients above 20--30 mmHg to identify meaningful residual or recurrent obstruction \cite{vanderpalen2025,kim2010lvoto}. Although the face-averaged CFD pressure difference is not numerically equivalent to a Doppler peak or catheter gradient, it directly measures the modeled pressure difference between the VSD and aortic-annulus planes and provides a consistent planning metric for comparing candidate designs. Our results demonstrate that we can reliably generate intracardiac baffle candidates with modeled pressure differences within clinically acceptable ranges.

The workflow shifts baffle design from a primarily intraoperative judgment to a patient-specific plan that can be inspected, revised, and tested preoperatively. Traditionally, surgeons fashion the patch while assessing the opened, arrested heart and estimating how the anatomy will change after the repair is pressurized. The computationally generated candidate is not intended to replace surgical judgment but rather provide an anatomically conformal reference whose modeled pressure difference can be assessed before surgery. Prospective studies should determine whether the use of prespecified patch templates can improve operative efficiency, including a reduction in aortic cross-clamp and cardiopulmonary bypass times, which are associated with adverse postoperative outcomes \cite{hasegawa_dependence_2005}.

Several limitations define the scope of this study. This four-patient cohort demonstrates technical feasibility but does not establish generalizability across the full anatomical spectrum of DORV. First, the workflow depends on the quality of the input segmentation. Machine learning-based segmentation methods were successfully applied to accelerate generation of patient-specific 3D models for surgical planning. The nnU-Net--based pipeline reduced total segmentation time to roughly 1--2 hours per case compared to approximately 4--8 hours of conventional manual segmentation \cite{pace2016}. This reduction in segmentation time is particularly important for prospective planning, where the interval between preoperative CTA acquisition and surgery may be limited. The most challenging regions for the model were near the VSD margins and aortic valve annulus, where reduced image contrast and spatial resolution can limit accuracy \cite{pouch2025}. All segmentations were therefore manually refined and reviewed by a cardiovascular imaging expert, retaining clinical quality control while using machine learning to accelerate the initial segmentation. Because independent reference segmentations were unavailable for this retrospective cohort, quantitative segmentation accuracy could not be evaluated. As additional cases are incorporated into an expanded dataset, model performance is expected to continue to improve, further reducing the need for manual editing and shortening overall segmentation time.

The current analyses evaluate one fixed surgeon-selected operative pathway. Although the selected attachment seam is reproducibly projected onto the anatomical mesh, its clinical placement remains surgeon-dependent. Related intra-atrial baffle studies have demonstrated both inter- and intra-surgeon variability in proposed patch pathways and hemodynamic differences between plausible patch designs \cite{nakamura2021patch,nakamura2025variability}. Different attachment seams may represent alternative operative strategies and could alter both the final baffle geometry and its modeled hemodynamic performance. Within the surgeon-selected attachment boundary, the method modifies and regularizes only the proposed baffle segment to meet the cross-sectional area target; it does not directly optimize curvature, eccentricity, local expansion, or streamline alignment. Future extensions can incorporate these features as additional design objectives and use CFD to compare alternative candidate geometries. Because the surrounding native right-ventricular myocardium is preserved, the method was evaluated only in anatomies where an LV-to-aorta pathway could be constructed through the native VSD. Patients requiring VSD enlargement were excluded, and the current framework does not model VSD enlargement or conal muscle resection. Future work can incorporate virtual VSD enlargement or conal resection before baffle construction, allowing the combined effects of myocardial resection and baffle geometry on postoperative hemodynamics. The framework can also vary the target area, $A_{\text{target}}$, to examine the trade-off between reducing the modeled LVOT pressure difference and encroachment on RV or tricuspid-valve space.

The current framework treats the baffle as a geometric construct and does not model the mechanics of the implanted pericardial patch. The smoothing step was used to avoid abrupt local changes in the proposed patch; however, the final implanted configuration will depend on patch stiffness, fixation treatment, and compliance mismatch with the surrounding tissue \cite{murdock2018, Aguiari2016, kaiser2021}. Furthermore, cardiac blood flow occurs within a deforming ventricular domain. Rigid-wall models do not fully capture intraventricular flow because ventricular contraction influences flow patterns throughout the cardiac cycle \cite{brown2024, tang2011}. Future work could incorporate FSI or coupled electromechanical and flow simulations with patient-specific ventricular kinematics and patch mechanics to better approximate the implanted baffle shape and outflow hemodynamics with cardiac wall motion.  Prior work on multiscale, multiphysics whole-heart modeling supports the feasibility of these approaches \cite{quateroni2017}.

The workflow is intended for preoperative planning and uses a normative, BSA-scaled flow condition because the postoperative flow state does not yet exist when the baffle is designed. The resulting CFD assessment therefore tests whether the proposed pathway can support the desired normal systemic flow, rather than predicting an individual patient's exact postoperative pressure gradient. In this retrospective cohort, matched postoperative baffle anatomy, systemic flow, and pressure measurements were not uniformly available. The postoperative Doppler LVOT velocities therefore provided only a contextual comparison. Continuous-wave Doppler reports the highest measured velocity component detected along the ultrasound beam, whereas the CFD metric represents the maximum pointwise velocity sampled along the medial centerline of the reconstructed flow domain \cite{quinones2002}. Targeted postoperative imaging and hemodynamic measurements in future cases will permit comparison of the planned and implanted pathways and the predicted and measured flow.

In conclusion, we developed a semi-automated computational framework for patient-specific design and evaluation of intraventricular baffles for biventricular repair. The framework combines patient-specific anatomy and a surgeon-selected baffle boundary with automated, area-constrained surface construction to generate watertight, CFD-ready VSD-to-aorta pathways. Applied retrospectively to four DORV cases, the workflow produced clinically reasonable geometries that maintained peak-flow pressure differences below the planning benchmark across mesh resolutions and under elevated flow. Although this study focused on DORV, the framework may support other forms of intraventricular baffling and biventricular outflow reconstruction, including biventricular conversion in borderline single-ventricle patients \cite{barnet2025}. Together, these results support prospective evaluation of the framework as a preoperative planning tool that allows surgeons to review and quantitatively assess candidate baffle geometries before surgery. The same approach can also be used to systematically compare repair strategies and support more objective, data-informed decision-making in these complex operations.

    \section{Declarations}
\label{sec:conclusions}

\subsection{Funding}
This work was supported by the National Science Foundation (DGE-2146755 to ESM and 2310909 to ALM). MRM and ALM were supported in part by a grant from the National Heart, Lung, and Blood Institute of the National Institutes of Health under award number R01HL173845. ADK was supported in part by a grant from the National Heart, Lung, and Blood Institute of the National Institutes of Health under award number K25HL175208, American Heart Association Grant 24CDA1272816 (\url{https://doi.org/10.58275/AHA.24CDA1272816.pc.gr.193564}), and a grant from the Stanford Maternal and Child Health Research Institute. AKR was supported in part by a grant from the National Heart, Lung, and Blood Institutes of the National Institutes of Health under award number R38HL143615.

\subsection{Competing interests}
The authors declare no competing interests.

\subsection{Author Contributions}
All authors contributed to the study conception and design. Methodology development was performed by Elena Sabdy Martinez and Alexander D. Kaiser. Data curation was conducted by Alexander K. Reed, Amit Sharir, and Shiraz A. Maskatia. Software development was performed by Elena Sabdy Martinez and Sascha W. Stocker. Validation was performed by Elena Sabdy Martinez, Alexander D. Kaiser, and Alexander K. Reed. Resources were provided by Shiraz A. Maskatia, Michael R. Ma, and Alison Lesley Marsden. 
The first draft of the manuscript was written by Elena Sabdy Martinez and Alexander K. Reed. All authors contributed to the reviewing and editing of the manuscript. Supervision was provided by Alexander D. Kaiser, Shiraz A. Maskatia, Michael R. Ma, and Alison Lesley Marsden. Project administration and funding acquisition were performed by Michael R. Ma and Alison Lesley Marsden. 

\subsection{Data availability}
All finalized meshes, including the partitioned surface models and SimVascular-compatible tetrahedral meshes for all four patients, are publicly available at: https://github.com/StanfordCBCL/Surgical-Baffle-Geometries.

    \newpage
    \bibliographystyle{unsrt}
    \bibliography{references}

\end{document}